# Magnetic phase diagram of interacting nanoparticle systems under the mean field model

**Zhongquan Mao and Xi Chen**[1]

Department of Physics, South China University of Technology, Guangzhou 510640, P. R. China

[1]E-mail: xichen@scut.edu.cn

Abstract

The disordered random-anisotropy magnetic nanoparticle systems with competing dipolar interactions and ferromagnetic exchange couplings are investigated by Monte Carlo simulations. Superspin glass (SSG) and superferromagnetic (SFM) behaviors are found at low temperatures depending on the interactions. Based on the mean field approximation, the Curie-Weiss temperature $T_{CW} = 0$ is suggested as the phase boundary between the SSG systems and the SFM systems, which is convinced by the spontaneous magnetizations and relaxations. The magnetic phase diagram is plotted.

## 1. Introduction

Disordered magnetic nanoparticle systems with competing interactions and random anisotropy have been the subject of intense interests [1]. It is now widely accepted that a low-temperature superspin glass (SSG) phase could exist besides the superparamagnetic (SPM) phase due to the dipolar interactions or RKKY interactions between particles [2-5]. Recently the superferromagnetic (SFM) state has been observed in the two-dimensional (2D) Co nanoparticles by magnetic force microscopy [6] as well as in CoFe discontinuous metal-insulator multilayers (DMIMs) by magneto-optical Kerr microscopy [7]. Theoretically the dipolar interactions cannot yield a ferromagnetic ground state in a disordered system [8]. Tunneling exchange coupling might play an important role in such metal-insulator granular films [7]. However these conjectures have still to be carefully checked by simulations [1].

Spin glasses are founded in the frustration and randomness of microscopic magnetic interactions. For a magnetic nanoparticle system, theoretical studies indicate that the random interactions and anisotropies would give rise to the SSG 'order' [9-10]. On the other hand, the ferromagnetic exchange couplings lead to the ferromagnetic order [11-13]. Hence for a random system with competing dipolar interactions and the exchange couplings, the competitions between SSG order and ferromagnetic order must yield rich magnetic properties in granular systems. Similar issues also occur widely in the amorphous rare earth-transition metal alloys [12, 13] and many colossal magnetoresistance (CMR) manganites [14, 15] However, most theoretic works include only one kind of the interparticle interaction, i.e. either dipolar interactions or exchange couplings, which interplay with the random anisotropy [9-13]. Few works have considered the competition of both interactions with random anisotropy and positions. In order to better understand such complex systems, we perform the Monte Carlo (MC) simulations on the disordered random-anisotropy systems with competing dipolar interactions and ferromagnetic exchange couplings. A main result we find is that the zero Curie-Weiss temperature*s* could be the phase boundary between the SSG systems and the SFM systems. Based on this mean field model, a magnetic phase diagram has been plotted. Since a SFM DMIM system also exhibits very similar behaviour found in the SSG system, such as memory and rejuvenation effect [1, 16], it is not easy to distinguish an SSG system and an SFM system experimentally, our result could provide a rough but simple way.

## 2. Simulation method

We consider $N = 512$ identical single-domain ferromagnetic nanospheres with volume $V$ placed in disordered positions without overlap in a $L \times L \times L$ cube. Hence the average distance between two nearest-neighboring particles is $a = L/8$. The magnetic moment of particle $i$ is assumed to be $M_S V \hat{s}_i$, where $M_S$ is the saturation magnetization and the unit vector $\hat{s}_i$ is the orientation. The direction of the easy axis of particle $i$ is denoted by $\hat{e}_i$ with the anisotropy constant $K$. The total energy reduced by the anisotropy energy $KV$ is written as

$$E = -\sum_i (\hat{s}_i \cdot \hat{e}_i)^2 + g \sum_{i,j} \frac{\hat{s}_i \cdot \hat{s}_j - 3(\hat{s}_i \cdot \hat{r}_{ij})(\hat{s}_j \cdot \hat{r}_{ij})}{r_{ij}^3} - J \sum_{r_{ij} \leq a} \hat{s}_i \cdot \hat{s}_j - h \sum_i \hat{s}_i \cdot \hat{H} \qquad (1)$$

where $g$, $J$ and $h$ are the reduced energies, namely the dipolar energy, exchange energy and Zeeman energy, respectively. It should be noted that the temperature $T$ in this paper is also reduced by $KV$. The external field direction $\hat{H}$ is along the $z$ direction, and $r_{ij}$ is the distance between particles $i$ and $j$ in the unit of $a$ ($\hat{r}_{ij}$ indicates direction of $r_{ij}$). It should be noted that $g = \frac{M_S^2}{K} \frac{V}{a^3}$, which is proportional to the particle concentration $V/a^3$, and $J$ is assumed to be independent of $r_{ij}$ within a radius of $a$ to simplify the questions. Periodic boundary conditions are considered in simulations. The Ewald summation method is used to calculate the long-range dipolar interactions, while a truncate technique with a cutoff radius of $a$ is used to calculate the short-range exchange couplings. Standard Metropolis algorithm [17] with local dynamics [10, 18] is applied to calculate the spin configurations, and the polar angle restriction proposed by Otero [10, 18] is adopted to update the orientation of the magnetic moments. All simulation curves displayed here were averaged over more than 150 independent samples with different initial conditions.

## 3. Results and discussions

The random and competing interparticle interactions of a magnetic nanoparticle ensemble may influence the dynamic properties in two different ways: (i) by affecting the barrier heights and hence the relaxation time of the individual particles and (ii) by giving rise to a collective behavior [9, 19]. Aging phenomena is a characteristic feature of the non-equilibrium dynamics of the collective phenomenon [9, 20, 21]. A straightforward way to establish aging in the present model is to calculate the autocorrelation function $C(t, t_w) = \frac{1}{N} \sum_i \langle \hat{s}_i(t_w) \cdot \hat{s}_i(t_w + t) \rangle$, where $\langle \cdots \rangle$ means an average over different realizations of the thermal noise and $t_w$ is the waiting time, measured from some quenching time [9]. Fig. 1 shows the autocorrelation functions of three typical systems, i.e. the noninteracting system, dipolar interacting system and exchange interacting system, for different waiting times at various temperatures respectively. It is clear that $C(t, t_w)$ for both dipolar and exchange interacting systems with intermediate strength show strong waiting-time dependence at low temperature. The longer the systems wait, the slower they decay. No aging effect can be observed in the noninteracting system and at high temperature where all systems are in SPM states. Noticed that many works have evidenced a SSG phase in the pure dipolar interacting nanoparticle systems [1, 9, 10] and a ferromagnetic order tends to be established in the pure exchange coupled systems [11]. These aging effects indicate undoubtedly the collective phenomena of the magnetic moments of the nanoparticles at low temperatures.

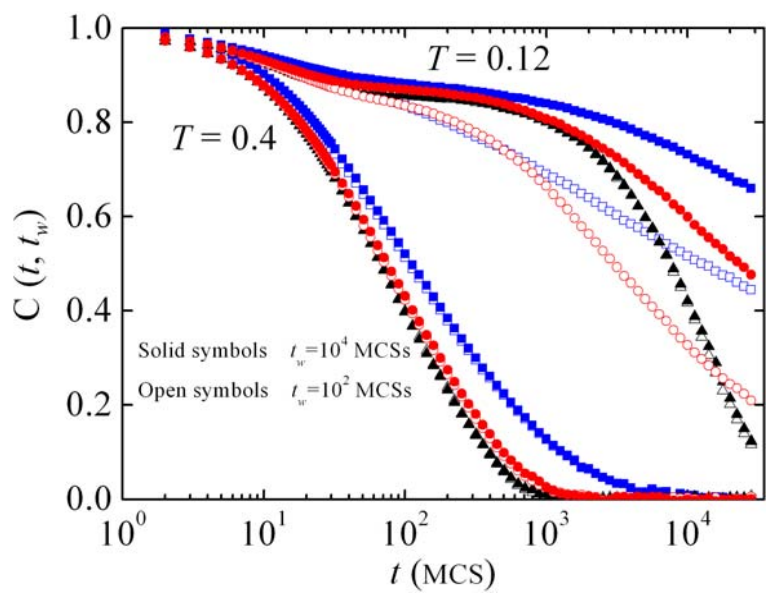

**Figure 1.** Autocorrelation function $C(t, t_w)$ as a function of MC steps $t$ for $t_w = 10^2$ (open symbols) and $10^4$ MCSs (solid symbols) at $T = 0.12$ and $T = 0.4$. The blue squares represent the $g = 0.2$, $J = 0$ system; the red circles represent the $g = 0$, $J = 0.2$ system; and the black triangles represent the noninteracting system. The applied field after waiting is $h = 0.02$ and the error bars are smaller than the symbols.

However, in a competing interactions system the aging effects cannot directly clarify the SSG and SFM state. To distinguish these two states, we consider the systems under a mean field approximation. Due to the random anisotropy, the mean anisotropy field of the system should be zero, which leads to a zero Curie-Weiss temperature, $T_{CW} = 0$. It is well known that the mean field of a ferromagnetic exchange coupling system is positive, which yields $T_{CW} > 0$. As for the dipolar interacting system, the first term of the dipolar energy, $g\dfrac{\hat{s}_i \cdot \hat{s}_j}{r_{ij}^3}$, is an antiferromagnetic coupling, which gives rise to a negative mean field. And the second term of the dipolar energy, $\dfrac{-3g(\hat{s}_i \cdot \hat{r}_{ij})(\hat{s}_j \cdot \hat{r}_{ij})}{r_{ij}^3}$, yields a random field which averages over the system should be zero. Thus for the disorder dipolar interacting systems, one should obtained $T_{CW} < 0$. [18, 22] Fig. 2 reveals the interaction-dependence of $T_{CW}$ in the interacting systems, which is in accord with the mean field predictions. In the competing interaction systems, $T_{CW}$ increases from negative to positive as $J$ increases, at fixed $g = 0.2$. For a small $J$ ($< 0.2$), $T_{CW} < 0$ indicates the dipolar interactions are dominant, where SSG may appear. For a larger $J$ ($> 0.2$), $T_{CW} > 0$ implies the exchange couplings win, where a ferromagnetic order tends to occur. Obviously, $T_{CW} = 0$, where $J = 0.2$, is the phase boundary between SSG and SFM state under the mean field approximation.

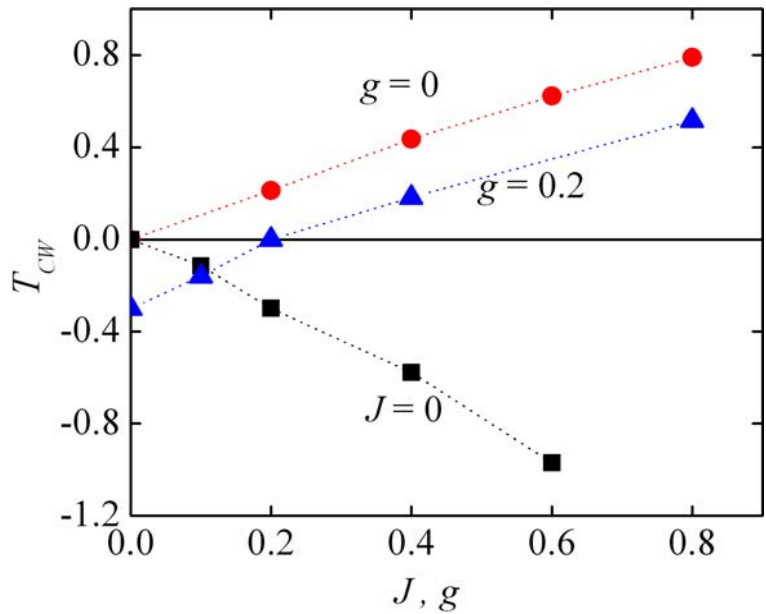

**Figure 2.** The Curie-Weiss temperature $T_{CW}$ for systems with different interactions. Here

$T_{CW}$ is obtained through extrapolating the linear part of the reverse ZFC/FC magnetization curves at high temperatures. $10^3$ MCSs were used for thermaliztion and the following $5\times10^3$ MCSs were used for average at each temperature. A small magnetic field $h = 0.05$ is applied during calculations. The error bars are smaller than the symbols.

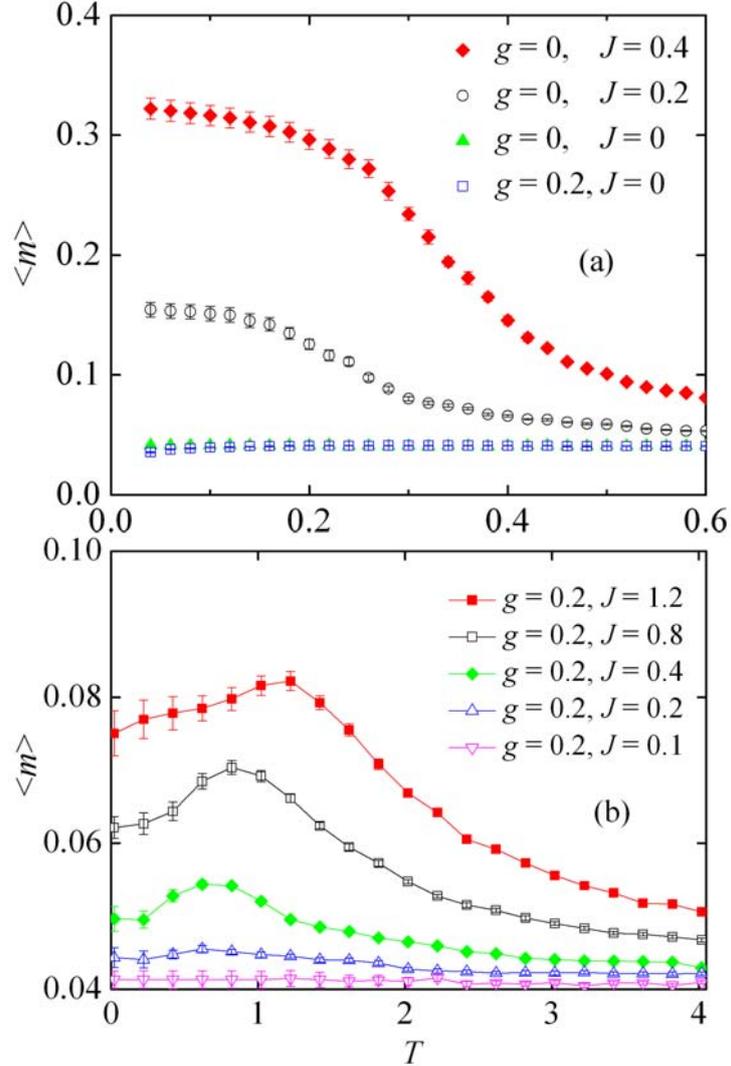

**Figure 3** The reduced spontaneous magnetization $<m>$ for: (a) noninteracting system and pure interacting (either dipolar interactions or exchange coupling) systems, and (b) competing interacting systems with fixed $g = 0.2$. $10^3$ MCSs were used for thermaliztion and the following $5\times10^3$ MCSs were used for average at each temperature.

The spontaneous magnetization is the order parameter of the ferromagnetic phase and it is always zero in spin glass phase. Fig. 3(a) reveals the reduced spontaneous magnetizations $<m>$ as the function of temperatures for noninteracting, dipolar interacting and exchange coupling systems respectively. Here the reduced spontaneous magnetization is given as $\langle m \rangle = \frac{1}{N}\left[\left(\sum_i s_{ix}\right)^2 + \left(\sum_i s_{iy}\right)^2 + \left(\sum_i s_{iz}\right)^2\right]^{1/2}$ during zero-field cooling. As expected, the exchange interacting systems exhibit significant spontaneous magnetizations <m> at low temperatures, while the noninteracting and the dipolar interacting systems show no distinct spontaneous magnetizations except for the background value due to the famous finite size effects [23]. Fig. 3(b) shows the spontaneous magnetizations <m> for the competing interaction systems with fixed g = 0.2. <m> decreases significantly comparing to the pure exchange coupling systems. Obviously the dipolar interactions suppress the long-range ferromagnetic orders. Remarkable spontaneous magnetizations <m> appear as $J > 0.2$ ($T_{CW} > 0$), which confirms the SFM orders. No distinct spontaneous magnetizations <m> are found in the systems with $J < 0.2$ ($T_{CW} < 0$). At the boundary for $J = 0.2$ ($T_{CW} = 0$), the spontaneous magnetizations are so unapparent that one cannot affirm the ferromagnetic order. This confirms the result from the mean field approximation.

As expected in a pure ferromagnetic exchange coupled system the spontaneous magnetization increases monotonically as the temperature decreases. However intermediate peaks are observed in Fig. 3(b) for $0.2 < J < 2.0$. Obviously, at low temperature the dipolar interactions interrupt the SFM orders of these systems, which lead to some glassy orders. Similar results were found in the dilute Heisenberg systems with competing ferromagnetic and antiferromagnetic exchange couplings, where 'reentrant spin-glass' was suggested. [24]

Another evidence for the mean field phase boundary comes from the simulations of the magnetic relaxations. It is found numerically and experimentally that the logarithmic magnetic relaxation rate for the nanoparticle systems decays by a universal power law after some crossover time $t_0$, [10, 25]

$$w(t) = -\frac{d}{dt}\ln m(t) = At^{-n}. \qquad (2)$$

Depending on the value of *n*, the relaxation function shows a stretched exponential decay (n < 1), or a power-law decay with a finite remanent $m_\infty$ (n > 1).

$$m(t) \approx \begin{cases} m_0 \exp\left[-(t/\tau)^{1-n}\right] & \text{for} \quad 0 \leq n < 1 \\ m_1 t^{-A} & \text{for} \quad n = 1 \\ m_\infty + m_1 t^{1-n} & \text{for} \quad n > 1 \end{cases}. \qquad (3)$$

Following the same process in Ref. [10], the magnetic relaxations for different competing interactions systems with fixed $g = 0.2$ are calculated. The fitting parameters are displayed in Table 1. For $J > 0.2$, $n > 1$, hence the finite remanents are expect according to Eq. (3), which convinces the SFM orders. For $J < 0.2$, $n \leq 1$ indicates SSG states. [1, 25]

Table 1. Temperature dependence of the exponent $n$ of the logarithmic magnetic relaxation rate (Eq. (2)) for systems with competing interactions at fixed $g = 0.2$.

| $J$ | $T_m$ | $n$ |
|---|---|---|
|  | 0.04 | 0.99 ± 0.01 |
| 0.1 | 0.16 | 0.98 ± 0.01 |
|  | 0.20 | 0.98 ± 0.02 |
|  | 0.04 | 1.00 ± 0.02 |
| 0.2 | 0.16 | 1.00 ± 0.02 |
|  | 0.20 | 1.02 ± 0.02 |
|  | 0.16 | 1.06 ± 0.01 |
| 0.4 | 0.24 | 1.05 ± 0.01 |
|  | 0.40 | 1.13 ± 0.03 |
|  | 0.16 | 1.17 ± 0.02 |
| 0.8 | 0.28 | 1.18 ± 0.03 |
|  | 0.40 | 1.17 ± 0.03 |

Based on the mean field model, the low temperature magnetic phase diagram is plotted in Fig. 4. Here the line of $T_{CW} = 0$ is used as the phase boundary between SSG and SFM order. For small particle concentration $V/a^3$, the phase boundary approximately follows the line $g = J$ where the exchange energies are comparative with the dipolar energies. For dense packed systems (e.g. for Co particles, $M_s = 1400$ Gs, $K = 10^6$ erg/cm$^3$, one can obtain the packing density $V/a^3 \approx 20\%$ for $g = 0.4$), the nanoparticles tend to align in order structure. [26] Less exchange couplings are required for SFM orders. Hence the phase boundary deviates from the line $g = J$.

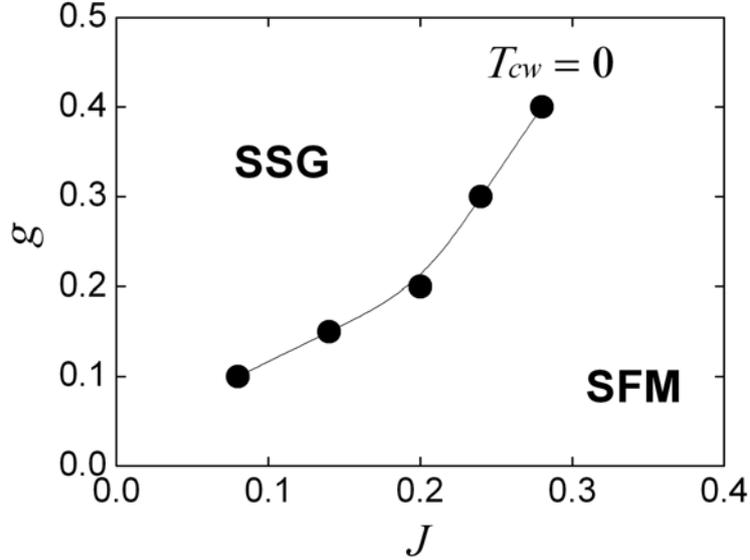

**Figure 4.** The low-temperature magnetic phase diagram of the dipolar and exchange competing nanoparticle systems.

The magnetic phase diagram has already plotted experimentally for the CoFe/Al$_2$O$_3$ DIMM systems. [25] The dipolar interactions competing with the tunneling exchanges between big particles via ultra small clusters are suggested. [7, 25] The SSG behaviors is found for nominal CoFe thicknesses $t_n < 1$ nm, where the dipolar interactions are dominated. At $t_n > 1.2$ nm, but below the percolation limit, $t_n = 1.8$ nm, the SFM domain state is encountered due to sufficiently strong exchange couplings. These results are qualitatively in agreement with our phase diagram. It is also found that at the crossover regime from SSG to SFM order, 1 nm $< t_n \leq 1.2$ nm, SFM states are observed at high temperature below $T_{CW}$ while spin glass-like behaviors occur at very low temperature. At these regimes, the tunneling exchange couplings might be stronger than the dipolar interactions at high temperature. Thus SFM orders occur. It is known that the tunneling exchanges decrease rapidly as the temperature decreases [27], while the dipolar interactions depend only on the spin configurations. Based on Fig. 4, if the tunneling exchange energy becomes smaller than the dipolar energy at low temperature, the system would reenter into a SSG state.

## 4. Conclusion

In conclusion, the disordered random-anisotropy system with competing dipolar interaction and ferromagnetic exchange coupling has been investigated by MC simulations. Due to the competing of the interactions, SSG states and SFM orders occur at low temperatures. The magnetic phase diagram is plotted based on a mean field approximation, where the line of $T_{CW} = 0$ is suggested to be the phase boundary between the SSG systems and the SFM systems. This is convinced by the spontaneous magnetizations and the relaxations. Since it is complicated to distinguish an SSG system and an SFM system experimentally, our results indicate that the $T_{CW} = 0$ could be a rough but simple criterion.

## Acknowledgements

Thanks are due to Prof. Dr. Wolfgang Kleemann for valuable discussions. This work is supported by the National Natural Science Foundation of China (No. 10704026), the Key Project of Chinese Ministry of Education (No. 109127), and the Fundamental Research Funds for the Central Universities, SCUT (No. 2009ZM0269).